\documentclass[sigconf,nonacm,pdfa]{acmart}
\usepackage[a-2b]{pdfx}
\usepackage[table]{xcolor}
\usepackage{xspace}
\usepackage{graphicx}
\usepackage{booktabs}
\usepackage{makecell}
\usepackage{multirow}
\usepackage{array}
\usepackage{tabularx}

\usepackage{amssymb}
\usepackage{amsfonts}
\usepackage{pifont}
\renewcommand{\checkmark}{\ding{51}}
\newcommand\vldbdoi{10.14778/3836663.3836720}
\newcommand\vldbpages{3718 - 3731}
\newcommand\vldbvolume{19}
\newcommand\vldbissue{11}
\newcommand\vldbyear{2026}
\newcommand\vldbauthors{\authors}
\newcommand\vldbtitle{\shorttitle}
\newcommand\vldbavailabilityurl{https://github.com/OpenDataBox/Dial}
\newcommand\vldbpagestyle{empty}

\newcommand{\hi}[1]{\noindent\textbf{#1}\xspace}
\newcommand{\llm}{\textsc{LLM}\xspace}
\newcommand{\llms}{\textsc{LLMs}\xspace}
\newcommand{\bfit}[1]{\textbf{\textit{#1}}}
\newcommand{\oursys}{\texttt{Dial}\xspace}

\newcommand{\zxh}[1]{#1}
\newcommand{\zl}[1]{#1}
\newcommand{\xhm}[1]{#1}

\newcommand{\zw}[1]{}
\newcommand{\yeye}[1]{}
\newcommand{\addcontent}[1]{}
\newcommand{\addcomment}[1]{}

\newcommand{\llabel}[1]{}
\newcommand{\RtwoColor}[1]{#1}
\newcommand{\RthreeColor}[1]{#1}
\newcommand{\RfiveColor}[1]{#1}
\newcommand{\RTwoMargin}[1]{}
\newcommand{\RThreeMargin}[1]{}
\newcommand{\RFiveMargin}[1]{}
\newcommand{\RTwoMarginTwo}[2]{}

\begin{document}

\title{Dial: A Knowledge-Grounded Dialect-Specific NL2SQL System}


\author{Xiang Zhang}
\authornote{Equal Contribution.}
\affiliation{%
  \institution{Shanghai Jiao Tong Univ.}
  \city{}
  \country{}
}
\email{zhangxxx@sjtu.edu.cn}

\author{Hongming Xu}
\authornotemark[1]
\affiliation{%
  \institution{Shanghai Jiao Tong Univ.}
  \city{}
  \country{}
}
\email{muzhihai@sjtu.edu.cn}

\author{Le Zhou}
\authornotemark[1]
\affiliation{%
  \institution{Shanghai Jiao Tong Univ.}
  \city{}
  \country{}
}
\email{mytruing1912@gmail.com}

\author{Wei Zhou}
\affiliation{%
  \institution{Shanghai Jiao Tong Univ.}
  \city{}
  \country{}
}
\email{weizhoudb@sjtu.edu.cn}

\author{Xuanhe Zhou}
\affiliation{%
  \institution{Shanghai Jiao Tong Univ.}
  \city{}
  \country{}
}
\email{zhouxuanhe@sjtu.edu.cn}

\author{Guoliang Li}
\affiliation{%
  \institution{Tsinghua University}
  \city{}
  \country{}
}
\email{liguoliang@tsinghua.edu.cn}

\author{Yuyu Luo}
\affiliation{%
  \institution{HKUST (GZ)}
  \city{}
  \country{}
}
\email{yuyuluo@hkust-gz.edu.cn}

\author{Changdong Liu}
\affiliation{%
  \institution{Shanghai Ideal Information Industry (Group)}
  \city{}
  \country{}
}
\email{liuzd4@telecom.cn}

\author{Guorun Chen}
\affiliation{%
  \institution{Shanghai Ideal Information Industry (Group)}
  \city{}
  \country{}
}
\email{chenguorun@telecom.cn}

\author{Jiang Liao}
\affiliation{%
  \institution{China Telecom Corporation Ltd. Shanghai Branch}
  \city{}
  \country{}
}
\email{liaojiang@telecom.cn}

\author{Fan Wu}
\affiliation{%
  \institution{Shanghai Jiao Tong Univ.}
  \city{}
  \country{}
}
\email{fwu@cs.sjtu.edu.cn}


\begin{abstract}
Enterprises commonly deploy heterogeneous database systems, each of which owns a distinct SQL dialect with different syntax rules, built-in functions, and execution constraints. However, most existing NL2SQL methods assume a single canonical dialect (e.g., SQLite) and struggle to produce queries that are both semantically correct and executable on target engines. Prompt-based approaches tightly couple intent reasoning with dialect syntax, rule-based translators often degrade native operators into generic constructs, and multi-dialect fine-tuning suffers from cross-dialect interference. 

In this paper, we present \oursys, a knowledge-grounded framework for dialect-specific NL2SQL. \oursys introduces: (1) a Dialect-Aware Logical Query Planning module that converts natural language into a dialect-aware logical query plan via operator-level intent decomposition and divergence-aware specification; (2) HINT-KB, a hierarchical intent-aware knowledge base that organizes dialect knowledge into $(i)$ a Canonical Syntax Reference, $(ii)$ a declarative function repository, and $(iii)$ a procedural constraint repository; and (3) an execution-driven debugging and semantic verification loop that separates syntactic recovery from logic auditing to prevent semantic drift. We construct DS-NL2SQL, a benchmark covering six major database systems with 2,218 dialect-specific test cases. Experimental results show that \oursys consistently improves translation accuracy by 10.25\% and dialect feature coverage by 15.77\% over state-of-the-art baselines. 

\end{abstract}


\maketitle

\pagestyle{\vldbpagestyle}
\begingroup\small\noindent\raggedright\textbf{PVLDB Reference Format:}\\
\vldbauthors. \vldbtitle. PVLDB, \vldbvolume(\vldbissue): \vldbpages, \vldbyear.\\
\href{https://doi.org/\vldbdoi}{doi:\vldbdoi}
\endgroup
\begingroup
\renewcommand\thefootnote{}\footnote{\noindent
This work is licensed under the Creative Commons BY-NC-ND 4.0 International License. Visit \url{https://creativecommons.org/licenses/by-nc-nd/4.0/} to view a copy of this license. For any use beyond those covered by this license, obtain permission by emailing \href{mailto:info@vldb.org}{info@vldb.org}. Copyright is held by the owner/author(s). Publication rights licensed to the VLDB Endowment. \\
\raggedright Proceedings of the VLDB Endowment, Vol. \vldbvolume, No. \vldbissue\ %
ISSN 2150-8097. \\
\href{https://doi.org/\vldbdoi}{doi:\vldbdoi} \\
}\addtocounter{footnote}{-1}\endgroup

\ifdefempty{\vldbavailabilityurl}{}{
\vspace{.3cm}
\begingroup\small\noindent\raggedright\textbf{PVLDB Artifact Availability:}\\
The source code, data, and/or other artifacts have been made available at \url{\vldbavailabilityurl}.
\endgroup
}


\section{Introduction}
\label{sec:intro}

\begin{sloppypar}
Existing NL2SQL methods predominantly target a single database dialect~\cite{DBLP:journals/tkde/LiuSLMJZFLTL25,DBLP:journals/corr/abs-2510-23587,DBLP:journals/pvldb/LiLCLT24, NIOWeaver}. However, in real-world scenarios, most enterprises (e.g., 80\% predicted by Gartner~\cite{technotes2025polyglot}) are supported by multiple database systems, each exposing its own SQL dialect with distinct syntax, function signatures, and compilation rules~\cite{empirical_microservices_2025,hr_tech_stack_2026,DMiM}. This creates the practical need for \textit{dialect-specific NL2SQL}: given a target database, the system must generate SQL that is both semantically correct and natively executable under that database's dialect. 









\begin{figure}[t] 
    \centering
    \includegraphics[width=.95\linewidth]{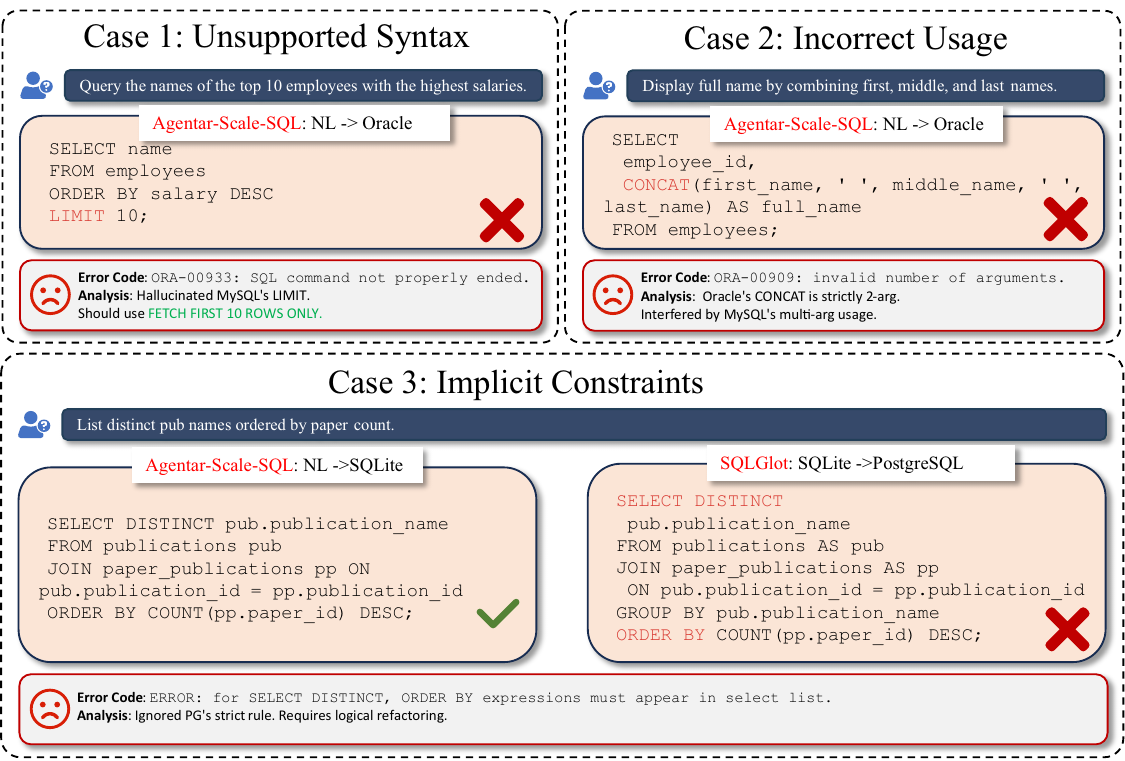} 
    \caption{Dialect-Specific NL2SQL Failures.} 
    \label{fig:motivation_badcases}
\end{figure}


Although recent enterprise-oriented benchmarks~\cite{Spider2.0,bird} have begun to incorporate different database dialects, their primary focus remains on schema complexity and cross-domain generalization. Consequently, current approaches often exhibit substantial performance degradation in dialect-specific NL2SQL: 

\noindent$\bullet$ \bfit{(1) Prompt-based methods} (e.g., DIN-SQL~\cite{DIN-SQL}, MAC-SQL~\cite{Wang2025@macsql}) rely on in-context demonstrations to guide generation. However, the underlying models are primarily pretrained and instruction-tuned on dominant dialects such as SQLite and MySQL. When deployed to a different database system, they often transfer familiar syntactic patterns, resulting in unsupported functions or incorrect operator signatures. For instance, in Figure~\ref{fig:motivation_badcases}, we can see \RFiveMargin{R5.D1}{\RfiveColor{\llabel{R5.D1.agentar-intro}Agentar-Scale-SQL~\cite{wang2025agentar}, a recent NL2SQL baseline that generates multiple SQL candidates and selects a final answer through a dedicated selection model,}} makes several typical dialect-relevant failures: 
$(i)$ In Case 1, Oracle does not support \texttt{LIMIT} and causes a parsing error; $(ii)$ In Case 2, Oracle's \texttt{CONCAT} accepts only two arguments, whereas MySQL accepts a variable number of arguments.

\noindent$\bullet$ \bfit{(2) Tool-augmented or rule-based pipelines} (e.g., SQLGlot~\cite{sqlglot}, WrenAI~\cite{wrenai}) typically rely on a generic intermediate representation to translate across dialects. While portable, this lowest-common-denominator strategy may overlook implicit dialect constraints, replace native operators with verbose rewrites, and fail in many cases (see Section~\ref{subsec:limit}). For instance, in Case 3 of Figure~\ref{fig:motivation_badcases}, PostgreSQL requires the \texttt{ORDER BY} expression in a \texttt{SELECT DISTINCT} query to appear in the projection list, a rule not imposed by SQLite. 

\noindent$\bullet$ \bfit{(3) Multi-dialect fine-tuning approaches} (e.g., ExeSQL~\cite{exesql}) attempt to internalize dialect variations within model parameters. However, different dialects share overlapping yet conflicting syntactic and functional patterns, which can easily induce cross-dialect interference and negative transfer~\cite{negativetransfer}. Moreover, this monolithic training paradigm lacks adaptability: supporting even a minor dialect update typically requires additional data for model finetuning.

\hi{Challenges.} There are three main challenges. 

\bfit{C1: How to map ambiguous intents to dialect-specific functions?} Users often express analytical requests in a dialect-agnostic manner (e.g., \textit{``months since registration''}), without specifying the concrete operators required to realize them. For example, computing month differences can correspond to \texttt{TIMESTAMPDIFF} in MySQL but require nested \texttt{EXTRACT}/\texttt{AGE} constructs in PostgreSQL. \RTwoMargin{R2.D2}{}\llabel{R2.D2.motivation}\RtwoColor{Although public documentation provides function definitions and syntax signatures, it is typically organized around isolated operators rather than intent-level realizations. Thus, the first challenge is not merely recalling documented syntax, but mapping an underspecified analytical intent to the correct dialect-specific realization.}

\bfit{C2: How to satisfy implicit dialect-level constraints?}
Even with the right function syntax, a query may still violate engine-specific parsing or semantic rules (e.g., \texttt{DISTINCT}-\texttt{ORDER BY} coupling, null-handling semantics). These constraints are rarely stated in user requests but directly determine executability. Hence, dialect-specific NL2SQL must ensure both intended functionality and dialect-level well-formedness. 

\bfit{C3: How to correct errors without semantic drift?}
Post-generation correction is unavoidable, but existing repair often targets executability through iterative re-generation. Such repair may overlook dialect nuances or alter the intended computation. Moreover, successful fixes are rarely consolidated into reusable experience, causing repeated reasoning for recurring dialect errors. Reliable dialect-specific NL2SQL therefore requires dialect-aware repair, semantic verification, and validated repair accumulation.





To address these challenges, we propose \oursys, a knowledge-grounded framework for dialect-specific NL2SQL. First, \oursys introduces a Natural Language Logical Query Plan (NL-LQP), which represents a user query as a dialect-agnostic
DAG of macro-operators capturing its core semantic intent. It then refines the plan into a dialect-aware form by identifying dialect-sensitive operators and aligning them with a standardized functional taxonomy. Second, \oursys constructs HINT-KB, a hierarchical intent-aware knowledge base that combines ANSI-based canonical syntax, dialect-specific function implementations, and engine-specific procedural constraints. Third, \oursys generates SQL through function-level retrieval from HINT-KB and iteratively repairs errors using execution feedback. Each executable query is further checked against the dialect-aware logical plan to prevent intent drift, and validated repair traces are distilled back into HINT-KB for continuous knowledge consolidation.

\hi{Contributions.} We make the following contributions.

\noindent(1) We propose \oursys, a knowledge-grounded dialect-specific NL2SQL framework that decouples intent modeling from dialect realization and achieves retraining-free executability.



\noindent(2) We design HINT-KB, an ANSI-grounded dialect knowledge base that separates function mappings from implicit compilation constraints and supports automated documentation distillation.

\noindent(3) We introduce NL-LQP, a dialect-agnostic DAG of
macro-operators that converts free-form user intent into structured relational operators and makes implicit computations explicit.

\noindent(4) We develop a divergence-aware dialect specification pipeline that detects dialect-sensitive operators and maps them to a standardized functional taxonomy for precise retrieval.

\noindent(5) We propose execution-driven refinement with syntactic recovery and semantic logic verification, preserving structural and computational invariants while ensuring executability.

\noindent(6) We construct DS-NL2SQL across six database systems. Compared with the strongest standard non-oracle baseline,
\oursys improves the mean per-dialect execution accuracy by
10.25\% and the mean dialect feature coverage by
15.77\%.

\end{sloppypar}

\section{Preliminary}
\label{sec:pre}

\begin{sloppypar}


\subsection{Problem Definition}
\label{subsec:def}

\hi{Dialect Specific NL2SQL.} Given a natural language question $q$, a database schema $\mathcal{S}$, and a target database dialect $d$, dialect-specific NL2SQL aims to generate a SQL query $s$ such that:
(1) $s$ is executable under dialect $d$, and
(2) $s$ correctly realizes the semantic intent of $q$ over $\mathcal{S}$. 
Compared with general NL2SQL task ~\cite{DBLP:journals/tkde/LiuSLMJZFLTL25, survey2, LLMDataSurvey, LLMPrepSurvey}, the output $s$ of dialect-specific NL2SQL is not a single canonical SQL form, but dialect-constrained realizations. 


\hi{Typical Dialect Discrepancies.} Based on our observations, dialect discrepancies mainly arise in three dimensions:




\noindent\underline{\emph{(1) Syntactic Rules:}} Database systems have distinct SQL syntax rules: (1) Identifier Quoting: Different characters are used to quote table or column names that might be reserved keywords (e.g., Oracle uses ``ID'', while MySQL uses `ID`). (2) Subquery Aliasing: Some databases like MySQL require all derived tables (subqueries in the \texttt{FROM} clause) have an alias, whereas others do not. (3) Pagination Syntax: The syntax for limiting the number of returned rows varies, such as \texttt{LIMIT} (MySQL/PostgreSQL) vs. \texttt{FETCH FIRST} (Oracle).

\noindent\underline{\emph{(2) Function Differences:}} The names and behaviors of built-in functions often vary: (1) String Manipulation: We can concatenate strings using the CONCAT() function, the || operator, or the + operator for different databases. (2) Date and Time Formatting: Functions that format dates and times have different names and argument styles (e.g., STRFTIME(), DATE\_FORMAT(), TO\_CHAR()).

\begin{figure*}[t]
\centering
\includegraphics[width=0.96\textwidth]{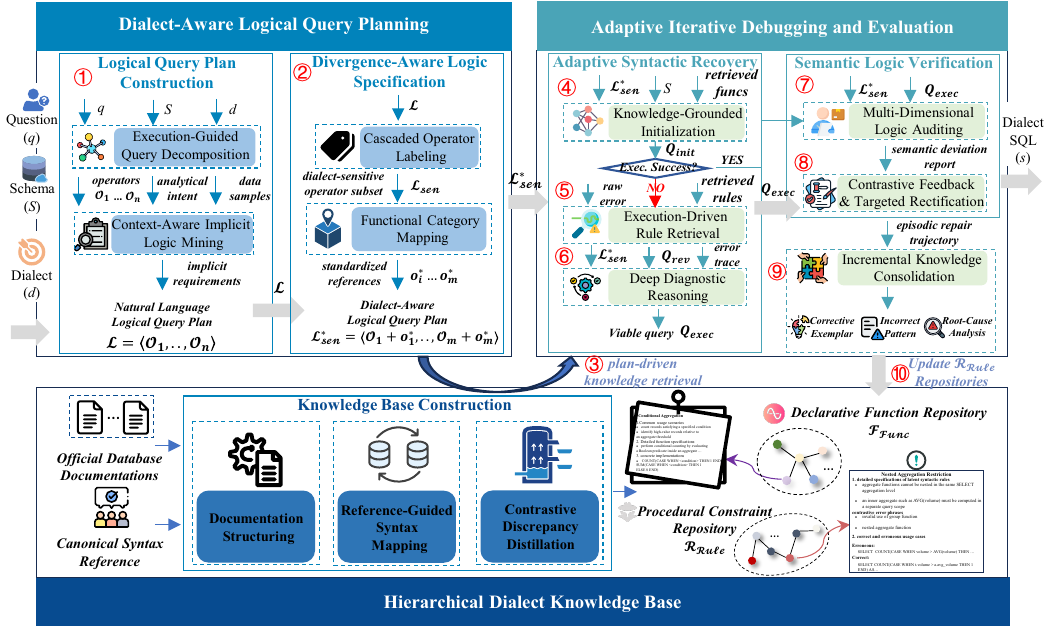} 
\caption{System Overview of \oursys.}
\label{DialectSQL}
\end{figure*}

\noindent\underline{\emph{(3) Semantic Variations:}} Implicit differences include:
(1) \emph{NULL value ordering}: some systems place \texttt{NULL} values first by default during sorting, while others place them last; moreover, explicit \texttt{NULLS FIRST} or \texttt{NULLS LAST} clauses are not supported in all databases. (2) \emph{Data type handling}: Core data types (e.g., \texttt{DATETIME}, \texttt{TIMESTAMP}, \texttt{DATE}) may differ in precision, representation, and functions.



\begin{figure}[t] 
    \centering
    \includegraphics[width=.95\linewidth]{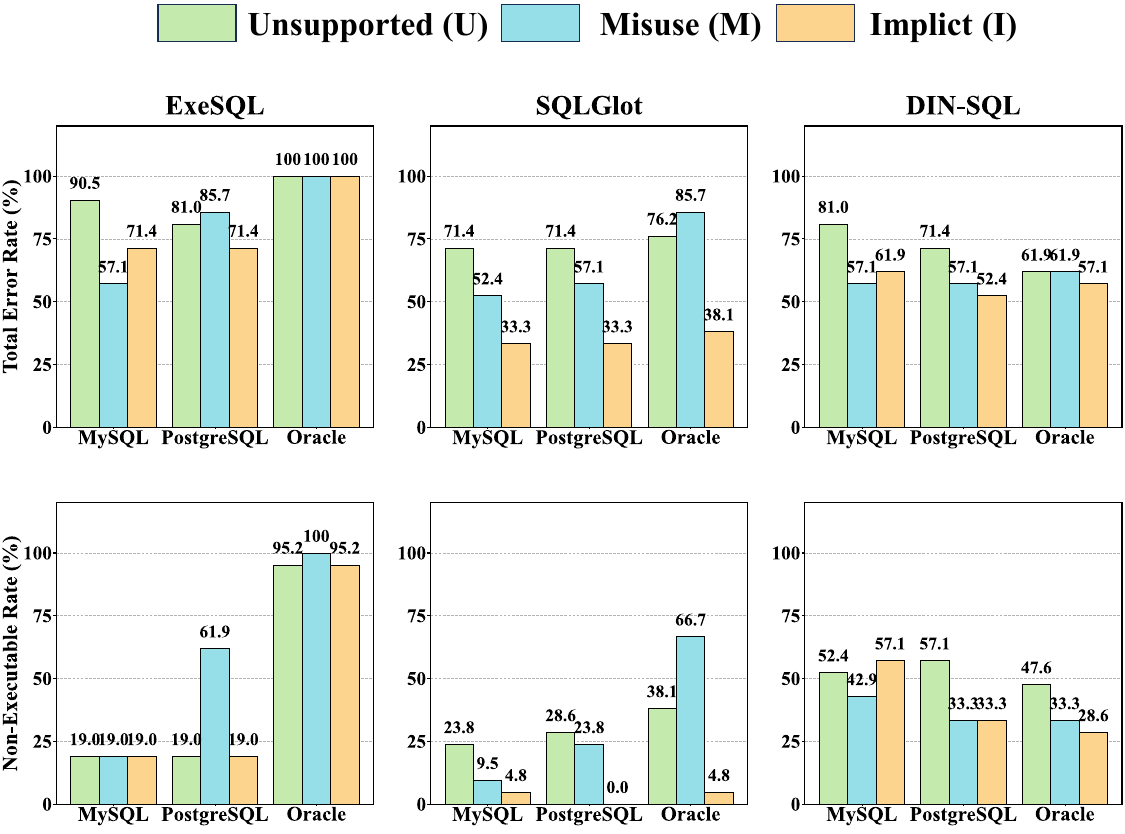} 
    \caption{Dialect-Specific Error Analysis.}
    \label{fig:preliminary_study}
\end{figure}


\subsection{Limitations of Existing Methods}
\label{subsec:limit}

We preliminarily evaluate ExeSQL as a fine-tuned model, SQLGlot as a rule-based transpiler, and DIN-SQL as a prompt-based method on three representative engines: MySQL, PostgreSQL, and Oracle. Figure~\ref{fig:preliminary_study} yields four observations.

\noindent\textbf{(Observation 1) Dialect-Specific Intent Realization.}
Existing methods often assume that once the user intent is understood, dialect-compliant SQL can be generated by prompting, fine-tuning, or rule-based rewriting. However, the same analytical intent may require different function signatures, argument conventions, and grammar forms across engines. Since such differences are rarely specified in user questions, static model parameters or handcrafted rules often fail to map abstract intent to valid dialect syntax, even when the high-level intent is correctly identified~\cite{DBLP:conf/emnlp/PourrezaR23}.

\noindent\textbf{(Observation 2) Missing Implicit Engine Constraints.}
Existing methods also struggle with implicit compilation constraints, such as dialect-specific dependencies between \texttt{DISTINCT}, \texttt{ORDER BY}, grouping, and \texttt{NULL} ordering. These constraints are not direct restatements of user intent, but engine-enforced rules that determine whether a query is legal. The high error rates on such features indicate that reliable dialect adaptation requires explicit modeling of implicit inter-clause and inter-functional constraints.

\noindent\textbf{(Observation 3) Dialect Overfitting from Fine-Tuning.}
Fine-tuning improves in-domain accuracy but can sharply reduce cross-engine generalizability. For example, ExeSQL with its released MySQL-tuned checkpoint (\texttt{exesql\_bird\_mysql}) shows 95\%--100\% non-executable rates on Oracle across the evaluated dialect features.

\noindent\textbf{(Observation 4) Executability Is Not Semantic Correctness.}
Executable SQL can still return wrong results. ExeSQL shows 38\%--62\% semantic drift on MySQL and PostgreSQL, while SQLGlot may produce syntactically valid but semantically mismatched translations under incomplete dialect mappings. Thus, dialect repair must verify logical fidelity, not merely restore executability.

\end{sloppypar}


\section{System Overview}
\label{sec:system_overview}

Figure ~\ref{DialectSQL} demonstrates the architecture and workflow of \oursys.

\hi{Dialect Knowledge Base Construction.} In the offline stage, \oursys builds a hierarchical dialect knowledge base (HINT-KB) from official documentation. Instead of using documentation as unstructured references, we reorganize it around a \textit{Canonical Syntax Reference}, which normalizes common database operations into an ANSI-aligned canonical space. 
HINT-KB is structured into \bfit{(1) Declarative Function Repository} mapping abstract functional intents (e.g., temporal arithmetic, string manipulation) to their concrete, dialect-specific implementations (e.g., \texttt{TIMESTAMPDIFF} in MySQL), which is indexed by natural-language usage patterns to enable direct intent-to-function retrieval; and  
\bfit{(2) Procedural Constraint Repository} capturing implicit structural rules required for execution correctness (e.g., quoting conventions), where these rules are indexed by diagnostic error signatures, enabling error-driven query correction during the debugging stage.\\\hi{Dialect-Aware Logical Query Planning.} In the online stage, given a user question $q$, schema $\mathcal{S}$, and target dialect $d$, \oursys first constructs a Natural Language Logical Query Plan $\mathcal{L}$ as a serialized DAG of macro-operators, which captures the dialect-agnostic analytical intent. It then identifies dialect-sensitive operators and annotates them with standardized functional categories from HINT-KB, producing the dialect-aware logical plan $\mathcal{L}_{sen}^{*}$. This plan serves as the shared intermediate representation for downstream knowledge retrieval, SQL generation, repair, and semantic verification.\\
\hi{Adaptive \& Iterative Debugging and Evaluation.} Based on the dialect-aware logical plan $\mathcal{L}_{sen}^{*}$ and retrieved function knowledge, \oursys enters a closed-loop generation and validation process. \bfit{(1) Knowledge-Grounded Initialization:}  
An initial candidate query is synthesized by retrieving dialect-specific function templates from the \textit{Declarative Function Repository} based on the labeled functional categories. 
\bfit{(2) Adaptive Syntactic Recovery:} If the query fails execution, the database error message is used as a key to retrieve a corresponding transformation rule from the \textit{Procedural Constraint Repository}, which is then applied to patch the query.
\bfit{(3) Semantic Logic Verification:} Once the query is executable, it is audited against the invariants defined in the original NL-LQP to prevent any semantic drift introduced during the repair process.
\bfit{(4) Incremental Knowledge Consolidation:} Finally, the verified repair logic is generalized into a new rule and integrated into HINT-KB, enabling \oursys to learn from its experience and continuously improve.

\section{Hierarchical Dialect Knowledge Base}
\label{sec:hint_kb}

To generate correct dialect-specific SQL, LLMs must be informed by two types of knowledge: (1) the explicit, intent-aligned functional syntax (e.g., using \texttt{TIMESTAMPDIFF} to calculate an age), and (2) the implicit, structural rules required for execution (e.g., quoting reserved keywords). However, naively retrieving from official documentation is inadequate, as these manuals are definition-oriented (e.g., technical details for \texttt{TIMESTAMPDIFF}) rather than intent-driven in user queries (e.g., calculating a user's age).

To address this, we propose the {Hierarchical Intent-aware Dialect Knowledge Base (HINT-KB)}, a structured repository that reorganizes vendor documentation into a queryable, dual-component architecture. \RTwoMargin{R2.W1, R2.D3}{\RtwoColor{\llabel{R2.W1.kb}\llabel{R2.D3.hintkb}This dual-part design guides downstream SQL generation with explicitly documented native implementations and execution-sensitive rules, rather than relying on dialect knowledge implicitly and unreliably encoded in LLM parameters.
}
} 

\subsection{Knowledge Base Architecture}
\label{sec:kb_design}





To resolve the semantic gap between explicit user intents and implicit database execution rules, the knowledge base HINT-KB is designed around two main principles: 

\noindent\textbf{Canonical Syntax Reference.} The Canonical Syntax Reference ($\mathcal{B}_{CSR}$) is \RThreeMargin{R3.W1}{\RthreeColor{\llabel{R3.W1.cold-start}a dialect-agnostic taxonomy of SQL functionality that serves as the initialization template for constructing HINT-KB for a new database.}} Since $n$ distinct user requirements (e.g., \textit{``sort by date''} or \textit{``get top 10''}) often map to a single functional requirement (e.g., \textit{result ordering and limitation}), which in turn requires $m$ atomic syntax points to implement (e.g., \texttt{ORDER BY} and \texttt{LIMIT} keywords), mapping syntaxes directly to specific requirements would result in an inefficient $O(n \times m)$ storage complexity. 
To eliminate redundancy, we group requirements with identical functional goals into unified categories, which serve as the fundamental units in HINT-KB. This optimization reduces the computational complexity to $O(m)$, effectively preventing knowledge base bloat. These categories are designed to be \textit{dialect-agnostic} and universally applicable across diverse database systems. Specifically, we abstract the comprehensive SQL syntax space into 11 distinct canonical categories (e.g., \textit{String Manipulation}, \textit{Date \& Time Operations}, and \textit{Window Functions}). This systematic categorization encompasses over 40 atomic syntax points. For instance, the category \textit{Date \& Time Operations} comprises 6 atomic syntax points, such as \textit{Date Truncation} (e.g., \texttt{DATE\_TRUNC}), \textit{Interval Arithmetic}, and \textit{Timestamp Extraction}.



\hi{Decoupled Retrieval Architecture.} 
HINT-KB employs a decoupled, dual-component architecture: the {Declarative Function Repository} ($\mathcal{F}_{Func}$), which is retrieved using the refined user intents (detailed in Section~\ref{sec:logical_planning}), and the {Procedural Constraint Repository} ($\mathcal{R}_{Rule}$), which is triggered by execution errors.

\noindent \RTwoMargin{R2.W2}\RThreeMargin{R3.D1}{}\llabel{R2.W2.ffunc}\llabel{R3.D1.ffunc-init}\RtwoColor{\bfit{(1) Declarative Function Repository ($\mathcal{F}_{Func}$)} stores dialect-specific implementations of functional constructs aligned with user intents. Each entry contains: (i) common usage scenarios (e.g., ``computing a person's age''); (ii) detailed function specifications that define the semantic operation (e.g., ``calculating the interval between two dates in years''); and (iii) concrete implementations that provide the exact syntactic realization in a target dialect (e.g., \texttt{TIMESTAMPDIFF(YEAR, ...)} in MySQL).}
These metadata elements are designed to be directly triggered by natural language requirements; for instance, the primitive $C_{\text{temporal\_diff}}$ can be invoked by the aforementioned age-related request. By providing such context-rich information, $\mathcal{F}_{Func}$ facilitates precise semantic disambiguation even when surface-level similarity is low. 

\noindent \bfit{(2) Procedural Constraint Repository ($\mathcal{R}_{Rule}$)} captures implicit structural rules exposed by execution errors. Each entry contains: (i) \textit{latent syntactic rule specifications} that define dialect-specific grammar constraints (e.g., reserved keywords such as \texttt{YEAR} cannot be used as unquoted aliases); and (ii) \textit{contrastive usage cases} that pair erroneous and corrected examples (e.g., \textit{Erroneous:} \texttt{SELECT ... AS YEAR} vs. \textit{Correct:} \texttt{SELECT ... AS "YEAR"}). \RTwoMargin{R2.W2}{}\RThreeMargin{R3.D2}
\RthreeColor{\llabel{R3.D2.rrule-evolution}}{\RtwoColor{\llabel{R2.W2.rrule}Rule specifications are normalized from documentation, while contrastive cases are dynamically evolved through execution-driven refinement (Section~\ref{subsec:ASR}). This separates documentation-grounded constraints from execution-grounded repair exemplars, enabling $\mathcal{R}_{Rule}$ to retain both static dialect rules and validated correction experience.}}

\subsection{Dialect Knowledge Base Construction}
\label{sec:kb_construction}

Manually populating the knowledge base for every target database is not scalable.
To automate this process, we designed a three-stage knowledge construction pipeline that systematically distills and organizes dialectal knowledge from raw, unstructured vendor manuals. 
This pipeline leverages the \textit{Canonical Syntax Reference} ($\mathcal{B}_{CSR}$) as a semantic bridge to overcome the mismatch between definition-oriented documentation and intent-driven translation tasks.

\hi{(1) Documentation Tagging.}
We first ingest the raw official documentation of a target dialect.
Documents in various formats (e.g., HTML, JSON, MD, SGML) are sequentially tagged by format-specific rule-based methods (e.g., \texttt{<dialect> txt <dialect>}). Their labeled contents are merged into a single document, yielding a processed version with clearly demarcated content sections.

\hi{(2) Reference-Guided Syntax Mapping.}
To populate the \textsf{HINT-KB} for a target dialect, this step maps the structured documentation to $\mathcal{B}_{CSR}$ via a dual-track semantic alignment. 
\RThreeMargin{R3.W1}{\RthreeColor{\llabel{R3.W1.csr-mapping}Rather than relying on ambiguous keyword matching, we perform retrieval by separately projecting canonical functional definitions and generic constraint templates from $\mathcal{B}_{CSR}$ into the documentation's vector space.
Specifically, canonical functional primitives serve as query inputs to identify semantically equivalent built-in functions; for example, the primitive of \textit{``extracting a substring''} is used to retrieve \texttt{SUBSTR} in \textsf{Oracle} or \texttt{CHARINDEX} in \textsf{SQL Server}, which are then organized into the Declarative Function Repository $\mathcal{F}_{Func}$.
Concurrently, generic constraint templates are utilized to locate corresponding structural rules within the manual; for instance, a generic template regarding \textit{``identifier quoting''} can successfully pinpoint the specific requirement in \textsf{PostgreSQL} documentation that \textit{``identifiers containing uppercase letters must be enclosed in double quotes''}, which is then normalized into the Procedural Constraint Repository $\mathcal{R}_{Rule}$.
Thus, $\mathcal{B}_{CSR}$ provides the functional primitives for constructing $\mathcal{F}_{Func}$ and the constraint templates for constructing $\mathcal{R}_{Rule}$, aligning vendor documentation into the two repositories of HINT-KB.}}

\hi{(3) Template-Based Knowledge Generation.}
\RThreeMargin{R3.D1}{\RthreeColor{\llabel{R3.D1.template-normalizer}After mapping official documentation to the predefined categories in $\mathcal{B}_{CSR}$, this module synthesizes structured knowledge entries from matched documentation fragments, with an LLM serving as a template normalizer. Specifically, for each functional requirement identified, the module generates a declarative function entry for $\mathcal{F}_{Func}$ by populating: (i) the \textit{usage scenario} inherited from the aligned canonical primitive (e.g., ``\textit{calculating age}''), (ii) the \textit{semantic specification} summarized from the matched documentation (e.g., ``\textit{date difference in years}''), and (iii) the \textit{concrete implementation} given by the target dialect documentation (e.g., \texttt{AGE()} in \textsf{PostgreSQL}).}}
For structural requirements, the module distills procedural constraints for $\mathcal{R}_{Rule}$ by defining the underlying grammar rules (e.g., \textit{``table aliases must use the AS keyword''}). Notably, at this stage, $\mathcal{R}_{Rule}$ entries consist only of these precise syntactic specifications; the contrastive erroneous/correct cases are not yet generated, as they are reserved for the dynamic execution-driven evolution phase (Section~\ref{subsec:ASR}). To supplement the \textit{Procedural Constraint Repository} ($\mathcal{R}_{Rule}$), this step targets documentation fragments that deviate from the ANSI baseline. Specifically, we design targeted prompts to instruct the LLM to identify dialect-specific constraints signaled by contrastive phrases, such as ``\textit{unlike standard SQL}'' or ``\textit{must be quoted as}.'' By filtering out extraneous descriptive text, the module isolates precise structural rules, such as reserved keyword conflicts. 
These structural deviations are inserted into $\mathcal{R}_{Rule}$ only as candidate constraints, while concrete erroneous/correct cases are supplied by the execution-driven evolution stage. They are promoted to trusted procedural repair knowledge only when later matched with real execution errors and validated by successful re-execution and semantic verification.
Crucially, we constrain the LLM with predefined templates, which is not allowed to invent new syntax, functions, or constraints beyond the retrieved documentation fragment. Its output must be strictly grounded in the cited documentation and normalized to the HINT-KB schema.


\section{Dialect-Aware Logical Query Planning}
\label{sec:dialect-extractor}

\begin{table*}[tb]
\centering
\caption{Standardized Logical Operators in NL-LQP.}
\label{tab:operators}
\footnotesize
\setlength{\tabcolsep}{3pt}



\renewcommand{\tabularxcolumn}[1]{m{#1}}

\newcolumntype{Y}{>{\raggedright\arraybackslash}X}
\newcolumntype{C}[1]{>{\centering\arraybackslash}m{#1}}

\begin{tabularx}{\textwidth}{
|C{0.17\textwidth}
|>{\hsize=1.5\hsize\linewidth=\hsize}Y
|>{\hsize=0.5\hsize\linewidth=\hsize}Y|}
\hline
\textbf{Operator}
& \multicolumn{1}{c|}{\textbf{Semantic Definition}}
& \multicolumn{1}{c|}{\textbf{Description Example}} \\
\hline

Data Sourcing ($\mathcal{O}_{src}$) 
& \RthreeColor{Specifies base relations and logical associations, including ANSI joins and dialect-specific correlations such as \texttt{LATERAL} or \texttt{APPLY}.}
& \RfiveColor{Source \texttt{Match} and \texttt{League} where \texttt{Match.league\_id = League.id}.} \\ 
\hline
Filtering ($\mathcal{O}_{flt}$) 
& Selects tuples using row-level or post-aggregation predicates.
& \RfiveColor{Keep orders after \texttt{2020-01-01} and with \texttt{status = 'completed'}} \\ 
\hline
Scalar Calculation ($\mathcal{O}_{cal}$) 
& Computes derived values through row-level transformations, such as casting, string operations, and temporal arithmetic.
& \RfiveColor{Compute \texttt{amount * 1.08} as \texttt{amount\_with\_tax}.} \\ 
\hline
Aggregation ($\mathcal{O}_{agg}$) 
& Changes data granularity by grouping records and computing summary metrics.
& \RfiveColor{Group by \texttt{customer\_id} and compute \texttt{AVG(amount)}.} \\ 
\hline
Result Organization ($\mathcal{O}_{org}$) 
& Organizes final results through projection, ordering, and cardinality control.
& \RfiveColor{Sort by \texttt{avg\_amount} descending and keep top 10.} \\ 
\hline
Auxiliary Operation ($\mathcal{O}_{aux}$) 
& \RfiveColor{Captures query-structuring requirements that organize, scope, or combine intermediate logical results, such as CTEs, nested queries, set operations, and window structures.}
& \RfiveColor{Use CTEs to compare department and company averages.} \\
\hline
\end{tabularx}
\end{table*}

\begin{figure*}[t] 
    \centering
    \includegraphics[width=.92\linewidth]{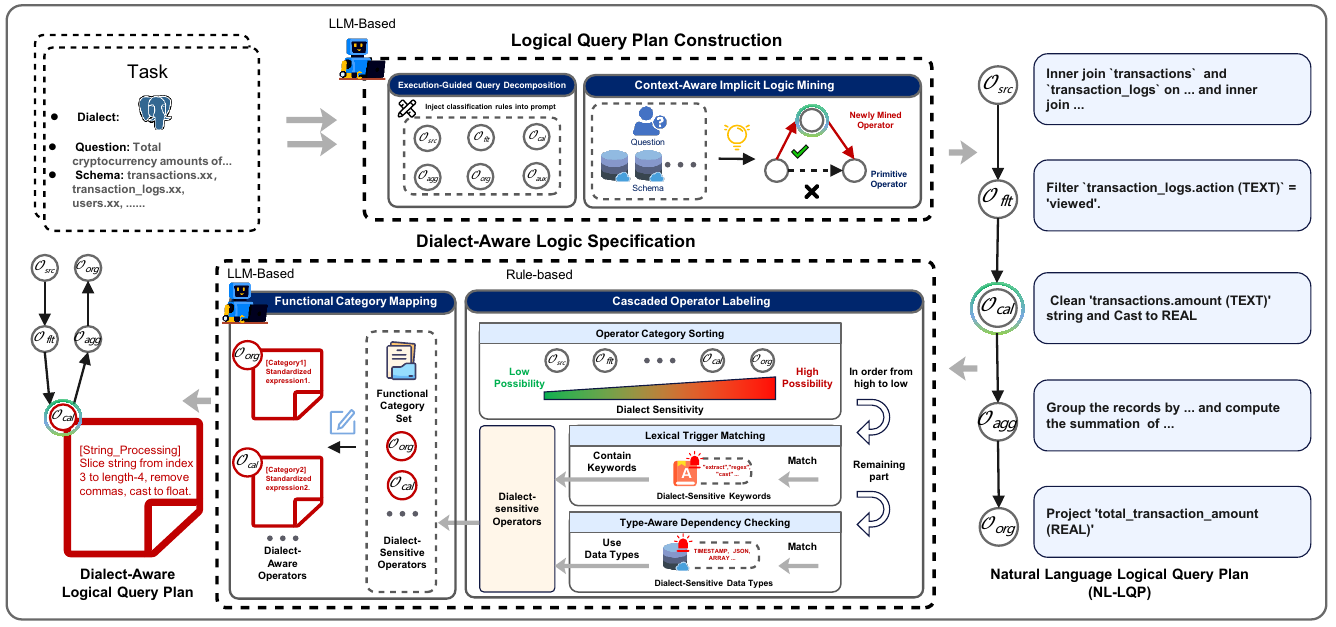} 
    \caption{Dialect-Aware Logical Query Planning.}
    \label{fig:Dialect-Aware Logical Query Planning}
\end{figure*}

Existing NL2SQL methods \cite{zhongSeq2SQL2017, DIN-SQL, Spider2.0, Wang2025@macsql} typically focus on mapping semantic user intent to SQL generation, often overlooking the usage of dialect-specific syntax.
It leads to a tight coupling between semantic understanding and dialect-specific constraints, resulting in errors arising from the neglect of dialect-specific syntax.
For instance, they might incorrectly invoke MySQL's \textsf{CHAR\_LENGTH} function in Oracle, especially with complex or implicit queries.
To address this challenge, we introduce the Dialect-Aware Logical Query Planning, which generates Natural Language Logical Query Plan (NL-LQP), as illustrated in Figure ~\ref{fig:Dialect-Aware Logical Query Planning}. \RTwoMargin{R2.W1\\R2.D3}{\RtwoColor{\llabel{R2.W1.planning}\llabel{R2.D3.planning}Since directly retrieving dialect knowledge from raw, ambiguous user requests introduces substantial noise, this module constructs a Natural Language Logical Query Plan (NL-LQP) to provide stable semantic anchors. It separates invariant query semantics from dialect-specific realization constraints, avoiding early mistakes on specific dialect syntax.}}
It operates in two phases:
(1) Constructing the semantic logical plan to decompose the query into operators (Section~\ref{sec:logical_planning}), and 
(2) Specifying the dialect-aware logic usages to tag potential dialect-sensitive elements (Section~\ref{sec:intent_anchoring}).

\subsection{Logical Query Plan Construction}
\label{sec:logical_planning}

To derive a strictly dialect-agnostic logical plan $\mathcal{L}$, we design an \llm-based structured generation method governed by strict constraints.
\RTwoMargin{R2.W2}{}\llabel{R2.W2.nl-lqp-output}\RtwoColor{Given the user query $q$, the target dialect $d$, and the database schema $\mathcal{S}$ (the DDL including data types and data samples), we generate an NL-LQP $\mathcal{L}$ that represents the query as a serialized operator DAG rather than a flat operator list. Each node corresponds to one instantiated macro-operator, and each edge records an explicit data-flow dependency between operators.}
We first present the standardized logical operators in the plan, and then describe the two-step construction.

\noindent \textbf{Standardized Logical Plan Operator.}
\RFiveMargin{R5.W1}\llabel{R5.W1.aux-text}{\RfiveColor{To isolate query semantics from dialect-specific syntax, we define a set of macro-operators as the basic units of NL-LQP. Each macro-operator represents one dialect-agnostic semantic step and is instantiated with four fields: an operator type, a natural-language operation description, an explicit index, and dependency references to preceding operators. As summarized in Table~\ref{tab:operators}, we use six operator types to cover the major roles in SQL query construction: relation sourcing, predicate filtering, scalar computation, aggregation, result organization, and auxiliary query structuring.
Most operators correspond to primary relational roles, while $\mathcal{O}_{aux}$ is used only for supplementary structures that organize complex queries but cannot be cleanly assigned to a single primary role, such as CTE scaffolding, nested-query decomposition, set operations, or window structures. For example, to ``find departments whose average salary is higher than the company-wide average'', the department-level and company-wide averages are computed by $\mathcal{O}_{agg}$, while the CTE or subquery structure that materializes and connects these intermediate results is represented by $\mathcal{O}_{aux}$. This design keeps NL-LQP expressive for complex queries while avoiding premature commitment to dialect-specific SQL syntax.}}

\noindent \textbf{Semantic-Driven Logical Plan Construction.}
\RTwoMargin{R2.D3.a}{}\llabel{R2.D3.a.twostep}\RtwoColor{Rather than asking an LLM to generate the entire plan in a single pass—which empirically causes the model to overlook implicit data transformations—we decompose this semantic construction into two explicitly constrained steps.}

\noindent \bfit{(1) Execution-Guided Query Decomposition.} 
To ensure the validity of the generated SQL queries, we require the \llm to construct the plan following a strict relational execution logic. 
This aligns the generated decomposition with how relational queries are logically processed (e.g., by first filtering raw data, then computing scalar metrics, and finally aggregating). \RTwoMargin{R2.W2}\RThreeMargin{R3.W7}\llabel{R3.W7.dag-structure}{\RtwoColor{\llabel{R2.W2.nl-lqp-dag}Crucially, the instantiated macro-operators form a Directed Acyclic Graph (DAG) connected by explicit data-flow dependencies. For LLM generation convenience, this DAG is serialized as a topologically ordered sequence. This sequential representation strictly dictates the logical decomposition order for SQL synthesis, rather than a physical execution plan, leaving structural optimizations (e.g., predicate pushdown) to the target database engine.}} To ground the logical operators in the physical database, we retrieve accurate schema metadata, including the exact data types and the data samples for relevant schema attributes.
The metadata is incorporated into the prompt context, ensuring that every referenced column explicitly carries its physical data type (e.g., \texttt{transactions.amount (TEXT)}).
Furthermore, to maintain strict semantic separation, the \llm is prohibited from outputting any SQL syntax or database-specific functions, and all operators should be expressed in natural language.
It finally projects the result with $\mathcal{O}_{org}$, outputting the username and the aggregated \texttt{total\_transaction\_amount}.

\noindent \bfit{(2) Context-Aware Implicit Logic Mining.}
To address the fact that users often omit necessary intermediate steps in their natural language queries, we introduce a heuristic compensation mechanism.
Without explicitly extracting these intermediate steps, SQL generation is likely to overlook essential functional requirements and introduce errors.
Specifically, we jointly analyze the user's analytical intent and the actual data samples embedded within the schema.
We identify implicit requirements that are not explicitly stated but are crucial for producing valid SQL.
For the same query in the last step, it asks for the ``total cryptocurrency amounts'' based on the \texttt{transactions.amount} column, the schema reveals that this column is of type \texttt{TEXT} (e.g., ``\$ 1,234.56 USD'').
The mechanism detects a conflict between the desired operation (summation) and the data type (string), which can lead to errors in SQL generation.
Since type conversion mechanisms differ across database dialects, the system explicitly handles this by materializing an implicit $\mathcal{O}_{cal}$ operator.
This operator cleans the string (e.g., stripping symbols and commas) and casts it into a numeric type.
The converted value is then passed to the $\mathcal{O}_{agg}$ operator for aggregation.

\subsection{Dialect-Aware Logic Specification}
\label{sec:intent_anchoring}

\begin{sloppypar}
\RTwoMargin{R2.W2}
\RFiveMargin{R5.D4}\llabel{R5.D4.underspecified}{\RtwoColor{\llabel{R2.W2.dialect-specification}Although the logical operators extracted in Section~\ref{sec:logical_planning} successfully capture the intended semantic steps, they remain semantically underspecified for direct dialect retrieval. For example, a scalar calculation operator like ``Convert \texttt{transactions.amount} (TEXT) into a numeric value'' correctly expresses the conversion intent, but lacks explicit sub-operations (e.g., removing currency symbols, stripping commas, and numeric casting).}} Using such free-form descriptions to retrieve syntactic rules from a dialect knowledge base would introduce substantial retrieval noise.
To augment the generic logical plan $\mathcal{L}$ for accurate downstream dialect-specific implementation, we introduce a \emph{Dialect-Aware Logic Specification} mechanism.
It links the logical operators to their corresponding dialect-specific syntax. Through two formalized sequential modules, we systematically augment the logical operators with standardized dialect specifications.
\end{sloppypar}

\noindent \textbf{(1) Cascaded Operator Labeling.} 
Given a generated plan with a sequence of logical operators $\mathcal{L} = \langle o_1, o_2, \dots, o_n \rangle$, we propose a rule-based cascaded labeling function, $F_{label}$, to efficiently isolate the subset of dialect-sensitive operators ($\mathcal{L}_{sen} \subseteq \mathcal{L}$).
This process minimizes computational overhead and reduces downstream retrieval noise.
Rather than relying on costly \llm inference, $F_{label}$ applies three hierarchical checks to evaluate each operator $o_i$: 
(i) \textit{Operator Category Sorting}:
$F_{label}$ first sorts operators with inherent structural divergence, such as scalar calculations ($\mathcal{O}_{cal}$) and result organization operations like cardinality constraints ($\mathcal{O}_{org}$) before those adhering to core ANSI SQL standards, like basic entity selection ($\mathcal{O}_{src}$) and simple equality filtering ($\mathcal{O}_{flt}$); 
(ii) \textit{Lexical Trigger Matching}:
To detect more complex operations hidden within simple operators, $F_{label}$ scans the descriptions of $o_i$ using a predefined lexicon of dialect-sensitive keywords (e.g., ``extract'', ``regex'', ``cast''); 
(iii) \textit{Type-Aware Dependency Checking}: Lastly, $F_{label}$ refers to the schema attributes with the schema definition to identify operators that manipulate complex or dialect-specific data types (e.g., TIMESTAMP, JSON, ARRAY), flagged as dialect-sensitive regardless of their operator category.
The operators identified by this cascaded pipeline form
$\mathcal{L}_{sen}$ and are forwarded to explicit dialect
augmentation.
\RTwoMargin{R2.D3.b}
{\RtwoColor{\llabel{R2.D3.b.rule-source}The rule base of $F_{\text{label}}$ is predefined offline: operator ordering comes from benchmark-level dialect-variation analysis and common SQL forms, lexical triggers are collected from recurring dialect-sensitive expressions and expanded by LLM-assisted paraphrases, and type dependencies are extracted from official documentation of the six target engines.}}

\begin{sloppypar}
\noindent \textbf{(2) Functional Category Mapping.} 
Since operators in $\mathcal{L}_{sen}$ have diverse textual descriptions, we define a functional category $\mathcal{C} = \{C_1, C_2, \dots, C_m\}$ to encode their functional characteristics into a unified semantic space. Each category $C_i$ represents a standardized category of dialect-specific operations (e.g., \texttt{[Temporal\_Manipulation]}, \texttt{[String\_Processing]}).
\RTwoMargin{R2.D3.b}{\RtwoColor{\llabel{R2.D3.b.standard-mapping}These categories are summarized from common dialect-specific functional patterns observed in mainstream benchmarks and vendor documentation are aligned with HINT-KB's functional taxonomy.}}
For each operator $o_i \in \mathcal{L}_{sen}$, we use an \llm as a semantic classifier and standardizer to map it to the corresponding category.
The \llm takes the verbose description of $o_i$ along with the category $\mathcal{C}$ as inputs, then assigns $o_i$ to the most relevant category $C \in \mathcal{C}$.
At the same time, the \llm is instructed to discard unnecessary explanations (e.g., business logic justifications) and focus on reformulating the core intent into a standardized textual format.
This results in a standardized reference $o^{*}_i = \langle C^*, \text{standard\_description} \rangle$.
For example, the description of \texttt{transactions.amount} is converted into a precise representation: $o^{*}_{cal} = \langle$ \texttt{[String\_Processing]}, ``Slice the amount string to remove currency prefix and unit suffix, remove commas, and cast the cleaned value to float'' $\rangle$. \RTwoMarginTwo{R2.W2}{R2.D3.b}{\RtwoColor{\llabel{R2.W2.dialect-aware-plan}\llabel{R2.D3.b.dialect-aware-plan}By appending these standardized references to the corresponding dialect-sensitive operators, \oursys transforms the generic plan $\mathcal{L}$ into the Dialect-Aware Logical Query Plan $\mathcal{L}_{sen}^{*}$. Specifically, $\mathcal{L}_{sen}^{*}$ preserves the original operator indices and dependency structure of $\mathcal{L}$, keeps dialect-insensitive operators unchanged, and augments only dialect-sensitive operators with a standardized functional category and a retrieval-oriented description.}}
These standardized representations then serve as high-precision semantic indices, guiding the downstream module to retrieve the appropriate syntactic patterns from the knowledge base. 
\end{sloppypar}
\RThreeMargin{R3.W2}{\RthreeColor{\llabel{R3.W2.plan-validation}To ensure the reliability of the generated plan, \oursys applies subsequent rule-based checks to validate the LLM outputs. Specifically, it verifies that each generated operator uses a predefined macro-operator type, references only schema-valid attributes, satisfies structural dependencies, and maps to a valid HINT-KB functional category. Outputs failing these checks are deterministically rejected and regenerated.}}

\section{Adaptive Debugging and Evaluation} 
\label{sec:AIDE}

\begin{sloppypar}
While HINT-KB provides foundational knowledge, it cannot guarantee one-shot executability due to LLM hallucinations and implicit dialect constraints discoverable only at runtime. For instance, an LLM might hallucinate a non-existent function like \texttt{DATE\_AGE()} for PostgreSQL instead of the correct \texttt{AGE()}, or overlook Oracle's prohibition of using column aliases in \texttt{WHERE} clauses. A naive refinement that directly feeds error messages back to the LLM often leads to \textit{semantic drift}, i.e., the model may alter the original business logic just to make the query run. To address this, we propose the Adaptive Debugging and Evaluation. \RTwoMargin{R2.W1\\R2.D3}{\RtwoColor{\llabel{R2.W1.aide}\llabel{R2.D3.debugging}Instead of relying on an unconstrained LLM self-repair loop, this module explicitly decouples syntactic error resolution from semantic logic auditing. By treating the dialect-aware plan ($\mathcal{L}_{sen}^{*}$) as an immutable ground truth, it ensures that execution-driven repairs resolve runtime compilation violations without inducing semantic drift.}} 

\subsection{Adaptive Syntactic Recovery}
\label{subsec:ASR}
This stage iteratively refines the generated query until successful execution is achieved. We collect diagnostic feedback from the target database and resolve detected violations through a structured three-phase recovery pipeline. This process ensures that the final query is not only syntactically well-formed but also compliant with dialect-level compilation constraints and executable.

\hi{(1) Knowledge-Grounded Generation.} \RTwoMargin{R2.D3.c}{}\llabel{R2.D3.c.init}\RtwoColor{The process begins by instructing an LLM to synthesize an initial candidate query, $Q_{init}$, using the dialect-aware plan $\mathcal{L}_{sen}^{*}$, the database schema $\mathcal{S}$, and relevant function templates from the Declarative Function Repository ($\mathcal{F}_{Func}$).} This query prioritizes functional mapping over structural compliance. \RTwoMargin{R2.D3.c}{}\llabel{R2.D3.c.error}\RtwoColor{If executing $Q_{init}$ fails due to dialect-specific syntactic violations (rather than general schema-linking errors like ``table not found''), the system captures the raw database error trace $\mathcal{E}_{raw}$ as a diagnostic signal.}

\hi{(2) Execution-Driven Rule Retrieval.} We then use the error trace $\mathcal{E}_{raw}$ and the associated failing SQL segment as a composite key to retrieve a specific transformation rule from the Procedural Constraint Repository ($\mathcal{R}_{Rule}$). This retrieval is driven by database feedback, ensuring the fix is targeted at the specific dialect violation. \RTwoMargin{R2.D3.c}{}\llabel{R2.D3.c.patch}\RtwoColor{The retrieved rule is then utilized to constrain an LLM to accurately patch $Q_{init}$, producing a revised query $Q_{rev}$.}

\hi{(3) Deep Diagnostic Reasoning.} If $Q_{rev}$ also fails, indicating a complex error not covered by existing rules, the system escalates to a deep diagnostic phase. \RTwoMargin{R2.D3.c}{}\llabel{R2.D3.c.deep}\RtwoColor{In this phase, an LLM acts as a diagnostic agent to perform a multi-dimensional root-cause analysis by cross-referencing the flawed query $Q_{rev}$, the new error trace, and the ground-truth intent preserved in $\mathcal{L}_{sen}^{*}$.} Here, $\mathcal{L}_{sen}^{*}$ serves as a crucial structural anchor, ensuring that necessary syntactic transformations do not inadvertently alter the core logic. \RTwoMargin{R2.W2}{}\llabel{R2.W2.deep-diagnostic}\RtwoColor{The diagnostic agent then localizes the error-triggering SQL fragment, generates candidate repairs, and accepts a candidate as $Q_{exec}$ only after successful execution and subsequent Semantic Logic Verification.}

\subsection{Semantic Logic Verification}
\label{subsec:semantic_verification}

\RFiveMargin{R5.W2}
\llabel{R5.W2.semantic-verification}
\RfiveColor{To ensure the executable query $Q_{exec}$ has not drifted from the user's intent, it undergoes an operator-level semantic verification procedure. Unlike approaches that rely on subjective self-reflection~\cite{correct2}, which can be prone to inconsistency without external grounding~\cite{correct1}, we leverage the dialect-aware plan $\mathcal{L}_{sen}^{*}$ as an objective gold standard.  Crucially, \oursys does not accept a query merely because it executes; it must pass the following semantic verifications.}

\noindent\RfiveColor{\hi{(1) Multi-Dimensional Logic Auditing.} We parse $Q_{exec}$ into an Abstract Syntax Tree (AST) and map its clauses to a sequence of logical operators. This allows for a normalized comparison against four semantic invariants derived from the macro-operators in $\mathcal{L}_{sen}^{*}$:
$(i)$ \textit{Structural Topology:} Verifies that the source relations and join dependencies in $Q_{exec}$ strictly match the data-sourcing operators ($\mathcal{O}_{src}$) defined in $\mathcal{L}_{sen}^{*}$.
 $(ii)$ \textit{Constraint Fidelity:} For each filtering operator ($\mathcal{O}_{flt}$) in $\mathcal{L}_{sen}^{*}$, \oursys extracts the required predicate specification, including referenced attributes, comparison relations, constants or threshold expressions, Boolean connectors, and expected SQL scope. It then verifies that these constraints are preserved in the corresponding AST scope (\texttt{WHERE}, \texttt{JOIN...ON}, or \texttt{HAVING}). A repair that drops a required predicate, changes a comparison direction, weakens a Boolean connector, or illegally moves an aggregate condition from \texttt{HAVING} to \texttt{WHERE} will be rejected as semantic drift.
$(iii)$ \textit{Computational Consistency:} Confirms that calculation and aggregation operators ($\mathcal{O}_{cal}, \mathcal{O}_{agg}$) preserve the intended computation, such as temporal arithmetic logic and grouping granularity, rather than substituting superficially executable but mathematically different expressions.
$(iv)$ \textit{Projection Accuracy:} Compares the output columns, aliases, ordering expressions, and cardinality constraints against the result-organization operator ($\mathcal{O}_{org}$) to ensure the returned structure remains consistent with the intended answer.}

\noindent\RfiveColor{\hi{(2) Contrastive Feedback and Rectification.} If any invariant is violated, \oursys generates a semantic deviation report of the mismatch. The report explicitly identifies the mismatched operator, the expected constraint from $\mathcal{L}_{sen}^{*}$, and the conflicting SQL fragment in $Q_{exec}$. This report is fed back to the reasoning module as a high-priority constraint for a targeted repair. For example, if $\mathcal{L}_{sen}^{*}$ specifies a recent-date predicate such as \texttt{order\_date >= CURRENT\_DATE - INTERVAL '30 days'}, a repaired query using the opposite comparison direction is rejected even if it executes, because the recovered $\mathcal{O}_{flt}$ violates the expected attribute, comparison relation, threshold, and \texttt{WHERE} scope. This cycle repeats until all invariants are satisfied, yielding the verified query $s$.}

\subsection{Incremental Knowledge Consolidation}
\label{subsec:knowledge_consolidation}
To eliminate redundant reasoning for recurring errors, the final stage distills successful repair trajectories into reusable knowledge for HINT-KB, enabling the knowledge base to evolve autonomously with execution-grounded repair experience.

\hi{(1) Knowledge Distillation.} A validated repair is abstracted into a generalized, schema-agnostic knowledge primitive, $\mathcal{G} = \langle P_{inc}, E_{cor}, A_{rtc} \rangle$. This structure formalizes the \textit{Incorrect Pattern} ($P_{inc}$), the \textit{Corrective Exemplar} ($E_{cor}$), and a natural-language \textit{Root-Cause Analysis} ($A_{rtc}$), transforming a one-off fix into a structured heuristic (e.g., mapping MySQL Error 1241 to its corresponding fix).

\hi{(2) Dual-Mechanism Knowledge Routing.} The new primitive $\mathcal{G}$ is then integrated back into HINT-KB. A routing decision is made based on the cosine similarity between $\mathcal{G}$ and the original logical plan $\mathcal{L}_{sen}^{*}$. 
\RThreeMargin{R3.W1}{\RthreeColor{\llabel{R3.W1.routing-enrichment}If the similarity is high ($\ge 0.75$), the fix is deemed intent-driven and is incorporated into the \textit{Declarative Function Repository} ($\mathcal{F}_{Func}$) as an intent-aligned functional realization. Otherwise, it is categorized as a universal, environment-driven constraint and routed to the \textit{Procedural Constraint Repository} ($\mathcal{R}_{Rule}$) as a contrastive erroneous/correct repair case.}}
\end{sloppypar}











\section{Experiments}
\label{sec:experiments}
In this section, we comprehensively evaluate \oursys across a heterogeneous environment comprising six major database systems: SQLite, PostgreSQL, MySQL, SQL Server, DuckDB, and Oracle.
We first detail the experimental setup, then present the construction of a novel, high-quality multi-dialect NL2SQL benchmark, and finally, we present a comprehensive analysis of the experimental results.

\begin{table*}[t]
    \centering
    \caption{Comparison of our Dialect-Specific NL2SQL Benchmark with other representative benchmarks.}
    \resizebox{\textwidth}{!}{
    \begin{tabular}{l c c c c c c c c}
        \toprule
        \textbf{Benchmark} & \makecell{\textbf{Dialect-Specific} \\ \textbf{NL2SQL}} & \makecell{\textbf{Multi-Dialect} \\ \textbf{NL-SQL}} & \makecell{\textbf{Dialectal} \\ \textbf{Incompatibility}} & \makecell{\textbf{Execution} \\ \textbf{Equivalence}} & \makecell{\textbf{\# Test} \\ \textbf{Samples}} & \makecell{\textbf{\# Dialect} \\ \textbf{Types}} & \makecell{\textbf{\# Test} \\ \textbf{Databases}} & \makecell{\textbf{Average Dialectal} \\ \textbf{Discrepancy}} \\
        \midrule
        Spider \cite{spider} & \ding{55} & \ding{55} & \ding{55} & \ding{55} & 2,147 & 1 & 206 & -- \\
        BIRD \cite{bird} & \ding{55} & \ding{55} & \ding{55} & \ding{55} & 1,789 & 1 & 15 & -- \\
        BIRD Mini-Dev \cite{bird} & \ding{51} & \ding{51} & \ding{51} & \ding{55} & 500 & 3 & 11 & 1.60 \\
        PARROT \cite{zhou2025parrot} & \ding{55} & \ding{55} & \ding{51} & \ding{51} & 598 & 8 & -- & -- \\
        Spider 2.0-Lite \cite{Spider2.0} & \ding{51} & \ding{55} & \ding{55} & \ding{55} & 547 & 3 & 158 & -- \\
        \midrule
        \textbf{DS-NL2SQL} & \textbf{\ding{51}} & \textbf{\ding{51}} & \textbf{\ding{51}} & \textbf{\ding{51}} & \textbf{2,218} & \textbf{6} & \textbf{796} & \textbf{3.67} \\
        \bottomrule
    \end{tabular}
    }
    \label{tab:benchmark_comparison}
\end{table*}

\subsection{Experimental Setup}
\label{subsec:setup}

\noindent\textbf{Baselines.} We compare \oursys against three categories of state-of-the-art approaches and use Qwen-3-Max as the default \llm backbone.
\textbf{(1) Input Prompting:} We evaluate {DIN-SQL}~\cite{DIN-SQL} and {Agentar-Scale-SQL}~\cite{wang2025agentar}. These methods rely primarily on \llms' in-context learning capabilities, driven by elaborate prompting strategies. Specifically, DIN-SQL uses Qwen-3-Max, while Agentar-Scale-SQL uses its released 32B generation model. Because its selection model is unavailable, we report both a standard single-candidate result and a top-8 oracle upper bound that selects the best of eight candidates; \textbf{(2) Model Finetuning:} These methods attempt to bridge the dialect gap by fine-tuning models on multi-dialect corpora.We evaluate
ExeSQL~\cite{exesql} with its released MySQL checkpoint.
SQL-GEN~\cite{sql-gen} is excluded because its weights are
unavailable;
\textbf{(3) Tool Augmentation:} These methods augment the generation process by integrating external parsers or rule-based translators. We evaluate Dialect-SQL~\cite{dialect-sql} and an Agentar-Scale-SQL+SQLGlot pipeline. Agentar first generates SQLite SQL, which SQLGlot~\cite{sqlglot} translates to the target dialect. Because SQLGlot receives LLM-generated SQL, it cannot repair upstream semantic errors. We omit WrenAI~\cite{wrenai} because it lacks SQLite support, connects to only one database at a time, and scales poorly to large schemas.

\noindent\textbf{Evaluation Metrics.}
We employ three metrics to measure performance:
\emph{(1) Executability (Exec):}
The percentage of generated SQL queries that execute without syntax errors on the target database;
\emph{(2) Execution Accuracy (Acc):}
The percentage of generated SQL queries that return result sets identical to the gold SQL;
\emph{(3) Dialect Feature Coverage (DFC):}
The recall of dialect-specific features (e.g., unique functions) successfully used in the generated SQL and present in the gold SQL.
Using predefined regular expression rules, it is a fine-grained metric that evaluates whether the methods use the intended database syntax correctly. 

\noindent\textbf{Implementation.}
All experiments and database executions are conducted on a workstation running Ubuntu 22.04 LTS, equipped with 512 GB of main memory and high-capacity storage. To ensure reproducibility and accurate execution feedback, we evaluate the generated queries against the following database versions: SQLite v3.45.3, MySQL v8.0.45, PostgreSQL v14.20, SQL Server v17.0, DuckDB v1.4.3, and Oracle Database 19c (Enterprise Edition).

\subsection{DS-NL2SQL Benchmark}
\label{ssec:bench}

Existing Text-to-SQL benchmarks, such as Spider~\cite{spider}
and BIRD~\cite{bird}, mainly use SQLite-compatible syntax and
do not capture the diversity of enterprise SQL dialects.
We therefore construct DS-NL2SQL, which contains 2,218 test
samples from 796 databases across six database systems.
As shown in Table~\ref{tab:benchmark_comparison}, it provides
parallel multi-dialect NL-SQL pairs and has an average dialectal
discrepancy of 3.67, compared with 1.60 for BIRD Mini-Dev.
We retain only queries that are incompatible with at least one
target system and manually verify execution equivalence across
dialects. To isolate dialect translation from schema linking, we
also provide the gold tables and columns as input. Thus, the
evaluated errors mainly reflect dialect syntax and engine
constraints. The construction pipeline has four steps:

\begin{table*}[t]
    \centering
    \setlength{\tabcolsep}{0.6mm} 
    \renewcommand{\arraystretch}{1.1} 

    \caption{Main performance comparison on DS-NL2SQL across six database dialects. \textbf{Exec}: Executability (\%), \textbf{Acc}: Execution Accuracy (\%), \textbf{DFC}: Dialect Feature Coverage (\%).}
    \label{tab:updated_results}
    
    \resizebox{\textwidth}{!}{%
    \begin{tabular}{l|ccc|ccc|ccc|ccc|ccc|ccc}
        \hline
        \multirow{2}{*}{\textbf{Method}} & \multicolumn{3}{c|}{\textbf{SQLite}} & \multicolumn{3}{c|}{\textbf{PostgreSQL}} & \multicolumn{3}{c|}{\textbf{MySQL}} & \multicolumn{3}{c|}{\textbf{SQL Server}} & \multicolumn{3}{c|}{\textbf{DuckDB}} & \multicolumn{3}{c}{\textbf{Oracle}} \\
        & Exec & Acc & DFC & Exec & Acc & DFC & Exec & Acc & DFC & Exec & Acc & DFC & Exec & Acc & DFC & Exec & Acc & DFC \\
        \hline
        
        \multicolumn{19}{c}{\textbf{\textit{Input Prompting}}} \\
        \hline
        DIN-SQL & 83.36 & 44.27 &  63.75 & 69.07 & 37.83 &  40.94 & 49.95 & 29.13 & 33.59 & 73.67 & 40.08 &  48.05 & 65.28 & 36.93 &  42.72 & 66.73 & 39.13 &  53.16 \\
        Agentar-Scale-SQL & 98.69 & 50.36 &  74.93 & 82.10 & 41.25 &  44.63 & 77.95 & 37.96 &  48.87 & 65.24 & 31.70 &  37.82 & 85.75 & 44.23 &  54.43 & 78.58 & 42.25 &  50.16 \\
        \RtwoColor{Agentar-Scale-SQL \textit{(top-8 upper bound)}} 
        & \textbf{99.86} & \textbf{61.90} & \RtwoColor{82.94} 
        & \RtwoColor{91.84} & \RtwoColor{50.45} & \RtwoColor{54.33} 
        & \RtwoColor{86.56} & \RtwoColor{46.26} & \RtwoColor{56.88} 
        & \RtwoColor{74.89} & \RtwoColor{41.07} & \RtwoColor{48.39} 
        & \RtwoColor{94.77} & \RtwoColor{48.30} & \RtwoColor{63.30} 
        & \RtwoColor{92.70} & \RtwoColor{48.76} & \RtwoColor{68.60}\\
        \hline

        \multicolumn{19}{c}{\textbf{\textit{Model Finetuning}}} \\
        \hline
        EXESQL & 86.88 & 26.96 & 44.03 & 80.12 & 26.65 &  28.29 & 84.36 & 26.69 &  37.07 & 54.37 & 18.26 &  20.05 & 81.24 & 27.23 &  35.12 & 5.50 & 3.74 & 4.25 \\
        \hline

        \multicolumn{19}{c}{\textbf{\textit{Tool Augmentation}}} \\
        \hline
        Dialect-SQL & 81.56 & 41.61 & 55.84 & 81.24 & 39.90 & 42.65 & 84.58 & 44.05 & 58.65 & 80.43 & 41.43 & 51.97 & 80.66 & 41.16 & 49.31 & 74.75 & 39.13 & 50.97 \\
        Agentar-Scale-SQL+SQLGlot & 98.60 & 50.36 & 74.93 & 81.51 & 42.61 &  70.72 & 90.17 & 47.57 &  67.89 & 80.79 & 42.52 &  68.33 & 82.78 & 43.15 &  65.54 & 81.97 & 43.55 &  58.43 \\
        \hline

        \oursys \textit{(Ours)} & 99.67 & 59.00 & \textbf{90.07} & \textbf{98.33} &\textbf{53.33} &  \textbf{78.70} & \textbf{99.87} &\textbf{55.87} &  \textbf{88.42} & \textbf{99.00} & \textbf{51.71} &  \textbf{85.70} & \textbf{99.93} & \textbf{57.94} &  \textbf{80.58} & \textbf{99.21} & \textbf{53.42} &  \textbf{76.97} \\
        \hline
    \end{tabular}%
    }
\end{table*}

\hi{(1) Data Aggregation and Context Decoupling.} To overcome the inherent sparsity of truly dialect-sensitive queries in individual benchmarks and construct a sufficiently large candidate pool, we aggregated data from multiple mainstream datasets, including Spider~\cite{spider}, BIRD~\cite{bird},
SparC~\cite{yu2019sparc}, CoSQL~\cite{yu2019cosql},
OmniSQL~\cite{li2025omnisql}, and
Archer~\cite{zheng-etal-2024-archer}.  For multi-turn conversational datasets like SparC and CoSQL, we employed LLMs to rewrite context-dependent queries into semantically complete, self-contained questions, eliminating contextual dependencies within dialogue turns.

\hi{(2) Dialect Migration and Syntax Validation.} Using SQLite \cite{SQLite} as the source dialect, we utilized SQLGlot \cite{sqlglot} to translate queries into five target dialects: MySQL \cite{MySQL}, PostgreSQL \cite{PostgreSQL}, SQL Server, DuckDB \cite{DuckDB}, and Oracle \cite{Oracle}.
\RThreeMargin{R3.W4}{\RthreeColor{\llabel{R3.W4.construction}SQLGlot was used only as an initial candidate translator, rather than as the final authority for ground-truth construction; all candidates were further checked by target-dialect parsing and execution validation.}}
We then enforced strict syntactic validation using parse trees of each target database, discarding samples with parsing errors to ensure syntactic viability. 

\hi{(3) Dialect Specificity Filtering.} \RThreeMargin{R3.W6}{\RthreeColor{\llabel{R3.W6.filtering}A critical step in our pipeline is ensuring the benchmark targets dialect nuances rather than standard ANSI-generic SQL (which predominantly constitutes existing benchmarks).}} We migrated the schemas to all target database systems using SQLAlchemy and executed the queries. If a query was compatible across \textit{all} systems (e.g., a simple \texttt{SELECT * FROM table}), it was deemed a "generic query" and excluded. We retained only queries that exhibited dialect exclusivity (i.e., failed on at least one system due to dialect mismatch).

\hi{(4) Consistency Verification and Manual Correction.} We verified the execution results across dialects to ensure logical equivalence. Furthermore, to address the limitations of automated tools (e.g., SQLGlot's failure to map SQLite's \texttt{GROUP\_CONCAT} to Oracle's \texttt{LISTAGG}), we performed meticulous manual corrections using official documentation. 
\RThreeMargin{R3.W4}{\RthreeColor{\llabel{R3.W4.manual-correction}These corrections inject dialect knowledge beyond SQLGlot's native translation rules, so the final gold SQL is human-verified rather than merely SQLGlot-translated.}}
This process resulted in a robust multi-dialect benchmark characterized by high heterogeneity.

\subsection{Performance Comparison}
\label{ssec:main_results}


Table \ref{tab:updated_results} presents the overall performance of \oursys and the baselines across six database dialects. 

\begin{figure}[t]
    \centering
    \includegraphics[width=\linewidth]{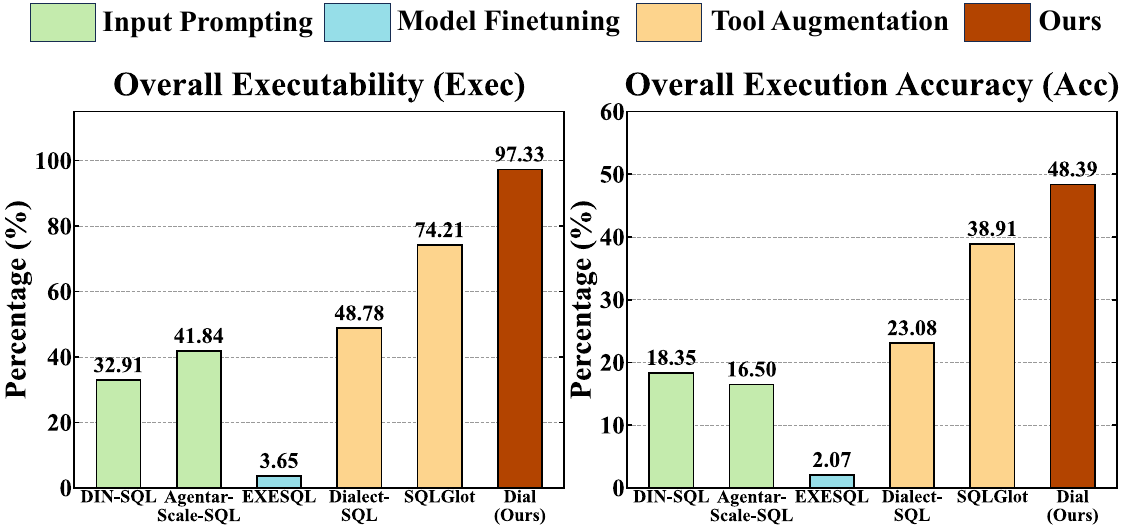}
    \caption{Overall Performance Comparison.}
    \label{fig:overall_performance}
\end{figure}

\hi{Translation Accuracy.}
Figure~\ref{fig:overall_performance} reports strict all-engine
metrics: a case is correct only if it executes (Exec) or returns
the gold result (Acc) on every evaluated engine. \oursys achieves
97.33\% overall Exec and 48.39\% Acc, improving over
Agentar-Scale-SQL+SQLGlot by 23.12 and 9.48 points,
respectively. While baselines might succeed on familiar dialects (e.g., SQLite), they suffer from single-point failures on complex dialects, causing their overall scores to plummet. For instance, model fine-tuning methods like EXESQL suffer from severe dialect overfitting; because it is statically fine-tuned on MySQL, it rigidly internalizes MySQL-specific syntax and fails catastrophically on Oracle (e.g., dropping to 5.50\% Exec), drastically dragging down its overall executability.

Despite its substantial gains over baselines, \oursys still fails in some cases, often due to uncontrollable factors in benchmark construction and evaluation: (1) \textbf{ambiguous natural-language specifications} where the question admits multiple reasonable SQL interpretations (e.g., “cumulative distribution” interpreted as running count vs. \texttt{CUME\_DIST()}); (2) \textbf{rigid output-format judging} where the evaluator accepts only a specific representation (e.g., \texttt{STRING\_AGG} vs. \texttt{ARRAY\_AGG}); and (3) \textbf{missing or undocumented benchmark-specific information} such as hard-coded IDs that cannot be derived from the schema or data samples. These factors are orthogonal to dialect-aware NL2SQL and would affect any system evaluated under the same benchmark.

\hi{Translation Robustness.} The performance gap is particularly pronounced in dialects with complex or highly unique syntax paradigms, such as Oracle and DuckDB. For instance, on Oracle, \oursys achieves 99.21\% Exec and 53.42\% Acc, whereas the Agentar-Scale-SQL+SQLGlot pipeline only reaches 81.97\% Exec and 43.55\% Acc. 
Tool-augmentation methods such as the Agentar-Scale-SQL+SQLGlot pipeline operate as deterministic transpilers over LLM-generated intermediate SQL rather than verified gold SQL; therefore, they inherit upstream NL2SQL errors and remain limited by uncovered dialect mappings in complex cases.
Meanwhile, Input Prompting methods (e.g., Agentar-Scale-SQL) suffer from severe ``dialect hallucinations'', mistakenly applying MySQL or PostgreSQL functions in Oracle. Even when evaluated under a top-8 oracle upper bound to fully exploit its test-time scaling potential, Agentar remains comparable to \oursys only on the widely-supported SQLite, while substantially lagging behind on the other five systems (e.g., lower by 5.16--24.11\% in Exec and 8.37--37.31\% in DFC). In contrast, \oursys utilizes a logic-decoupled architecture to accurately isolate user intents before anchoring them to dialect-specific implementations.

\textbf{Latency vs. Accuracy Trade-off.} \oursys achieves a favorable latency--robustness tradeoff. We conduct a separate latency experiment on 300 cases. \oursys takes 28.01 seconds per case on average, comparable to the Agentar-Scale-SQL top-8 upper bound at 27.46s, while achieving much higher Exec (99.67\% vs. 84.00\%) and Acc (43.00\% vs. 37.33\%). The latency breakdown shows that NL-LQP generation takes 9.33s, Dialect-Aware LQP tagging takes 5.54s, $\mathcal{F}_{Func}$ retrieval takes 4.15s, and SQL translation with feedback iteration takes 8.99s on average. Furthermore, \oursys provides a configurable pipeline for latency-sensitive deployments: disabling the feedback iteration reduces the average latency to 19.02s while still maintaining a highly competitive 89.67\% execution rate, demonstrating a flexible balance between real-time responsiveness and rigorous dialect correctness.

\subsection{Fine-Grained Analysis}
\label{ssec:fine_grained}

\begin{table*}[t]
\centering
\caption{Performance Comparison of \oursys Variants.}
\label{tab:ablation_updated}
\resizebox{\textwidth}{!}{
\begin{tabular}{c|ccc|ccc|ccc|ccc} 
\toprule
\multirow{2}{*}{\textbf{Method}} & \multirow{2}{*}{\makecell{\textbf{Logic Query} \\ \textbf{Planning}}} & \multirow{2}{*}{\textbf{HINT-KB}} & \multirow{2}{*}{\textbf{Correction}} & \multicolumn{3}{c|}{\textbf{PostgreSQL}} & \multicolumn{3}{c|}{\textbf{SQL Server}} & \multicolumn{3}{c}{\textbf{Oracle}} \\
\cline{5-13}
 & & & & Exec & Acc & DFC & Exec & Acc & DFC & Exec & Acc & DFC \\
\midrule
\textbf{\oursys} & \ding{51} & \ding{51} & \ding{51} & \textbf{98.33} & \textbf{53.33} & \textbf{78.70} & \textbf{99.00} & \textbf{51.71} & \textbf{85.70} & \textbf{99.21} & \textbf{53.42} & \textbf{76.97} \\ 
\midrule
\multirow{5}{*}{\textbf{Ablation Variants}} 
 & \RtwoColor{1-step NL-LQP} & \ding{51} & \ding{51} 
 & \RtwoColor{95.52} & \RtwoColor{49.04} & \RtwoColor{68.15} 
 & \RtwoColor{91.48} & \RtwoColor{47.09} & \RtwoColor{69.49} 
 & \RtwoColor{93.72} & \RtwoColor{50.73} & \RtwoColor{66.51} \\ 
 & \ding{55} & \ding{51} & \ding{51} & 91.64 & 38.59 & 53.35 & 98.16 & 48.35 & 69.13 & 95.16 & 43.99 & 52.41 \\ 
 & \ding{51} & \ding{51} & \ding{55} & 96.98 & 43.17 & 58.70 & 93.21 & 44.49 & 61.90 & 98.35 & 43.76 & 51.97 \\ 
 & \ding{51} & \ding{55} & \ding{55} & 96.53 & 45.94 & 61.92 & 95.63 & 46.71 & 61.26 & 91.34 & 43.15 & 48.52 \\ 
 & \ding{55} & \ding{51} & \ding{55} & 91.86 & 38.56 & 53.35 & 83.30 & 44.05 & 58.39 & 95.40 & 43.07 & 52.41 \\ 
\bottomrule
\end{tabular}
}
\end{table*}

\begin{table*}[!t]
\centering
\caption{Dialect-Specific Generation Errors Effectively Resolved by \oursys
(User questions are abstracted as realistic questions. Target syntax highlights the correct dialect-specific pattern alongside strictly prohibited anti-patterns).}
\label{tab:dialect_errors}
\resizebox{\textwidth}{!}{
\begin{tabular}{|c|c|c|c|c|c|c|c|c|}
\hline
\rowcolor{black}
\textcolor{white}{\textbf{Type}} & \textcolor{white}{\textbf{User Question (Abbreviated Intent)}} & \textcolor{white}{\textbf{Target Syntax Constraints (Gold)}} & \textcolor{white}{\textbf{DIN-SQL}} & \textcolor{white}{\textbf{EXESQL}} & \textcolor{white}{\textbf{Dialect-SQL}} & \textcolor{white}{\textbf{Agentar+SQLGlot}} & \textcolor{white}{\textbf{\oursys}} & \textcolor{white}{\textbf{Target Dialect}} \\
\hline
U1 & List all IP addresses accessed by each city. & \textbf{LISTAGG(...)} (Strictly NO \texttt{GROUP\_CONCAT}) & $\times$ & $\times$ & $\times$ & $\times$ & \checkmark & Oracle \\ \hline
U2 & Find the shipment details matching the ID. & \textbf{CAST(id AS CHAR)} (Strictly NO \texttt{AS TEXT}) & $\times$ & \checkmark & \checkmark & $\times$ & \checkmark & MySQL \\ \hline
U3 & When did the earliest complaint start on 2017-03-22? & \textbf{TO\_DATE(...)} (Strictly NO \texttt{date()} function) & $\times$ & $\times$ & $\times$ & $\times$ & \checkmark & Oracle \\ \hline
U4 & What are the total sales metrics for last month? & \textbf{Native Numeric} (Strictly NO \texttt{ORM bindings}) & \checkmark & \checkmark & $\times$ & \checkmark & \checkmark & DuckDB \\ \hline
M1 & Combine first, middle, and last names. & \textbf{first || middle || last} (Strictly NO 3-arg \texttt{CONCAT}) & $\times$ & $\times$ & $\times$ & \checkmark & \checkmark & Oracle \\ \hline
M2 & Retrieve user details across multiple related logs. & \textbf{FROM "T1" "T2"} (Strictly NO \texttt{AS} keyword) & \checkmark & $\times$ & \checkmark & \checkmark & \checkmark & Oracle \\ \hline
M3 & Sort the movie ratings from lowest to highest. & \textbf{ORDER BY rating ASC} (Strictly NO \texttt{asc()}) & \checkmark & $\times$ & $\times$ & \checkmark & \checkmark & Oracle \\ \hline
M4 & Find songs where the language is exactly English. & \textbf{col = 'english'} (Strictly NO double quotes \texttt{" "}) & \checkmark & \checkmark & \checkmark & $\times$ & \checkmark & Postgres/MySQL \\ \hline
M5 & What is the total number of districts? & \textbf{SELECT ... FROM DUAL} (Strictly NO absent \texttt{FROM}) & \checkmark & $\times$ & $\times$ & $\times$ & \checkmark & Oracle \\ \hline
M6 & Find the lowest stock product among top sellers. & \textbf{FROM (SELECT...) AS alias} (Strictly NO anonymous) & $\times$ & \checkmark & \checkmark & $\times$ & \checkmark & Postgres/MySQL \\ \hline
I1 & What are the names and total distinct programs used? & \textbf{GROUP BY...} (Strictly NO \texttt{DISTINCT} inside \texttt{OVER()}) & $\times$ & \checkmark & \checkmark & $\times$ & \checkmark & Postgres/MySQL \\ \hline
I2 & Count days with high trading volume. & \textbf{COUNT(CASE WHEN...)} (Strictly NO nested \texttt{AVG} aggregation) & \checkmark & \checkmark & $\times$ & \checkmark & \checkmark & MySQL \\ \hline
I3 & Find the highest revenue movie and its average rating. & \textbf{Isolate in CTE} (Strictly NO unaggregated \texttt{ORDER BY}) & $\times$ & $\times$ & $\times$ & \checkmark & \checkmark & SQL Server \\ \hline
I4 & What was the average price of the most recent crypto? & \textbf{= (SELECT... LIMIT 1)} (Strictly NO scalar unconstrained) & \checkmark & \checkmark & \checkmark & $\times$ & \checkmark & Postgres/MySQL \\ \hline
\end{tabular}
}
\end{table*}

\begin{table}[!t]
\centering
\caption{Performance over Different \llm Backbones.}
\label{tab:llm_comparison}
\resizebox{\columnwidth}{!}{
\begin{tabular}{l|ccc|ccc|ccc} 
\toprule
\multirow{2}{*}{\textbf{Method}} & \multicolumn{3}{c|}{\textbf{PostgreSQL}} & \multicolumn{3}{c|}{\textbf{SQL Server}} & \multicolumn{3}{c}{\textbf{Oracle}} \\
 & Exec & Acc & DFC & Exec & Acc & DFC & Exec & Acc & DFC \\ 
\midrule
\textbf{\oursys (Ours)} & & & & & & & & & \\
\quad Qwen-3-Max & \textbf{98.33} & \textbf{53.33} & \textbf{78.70} & 99.00 & \textbf{51.71} & \textbf{85.70} & 99.21 & \textbf{53.42} & \textbf{76.97} \\ 
\quad DeepSeek-V3.2 & 96.42 & 43.10 & 63.10 & 99.42 & 45.39 & 65.27 & 98.03 & 45.77 & 60.08 \\
\quad GPT-5.2 & 96.88 & 48.91 & 70.95 & \textbf{99.76} & 51.68 & 76.84 & \textbf{99.70} & 48.89 & 66.64 \\
\midrule
\textbf{DIN-SQL} & & & & & & & & & \\
\quad Qwen-3-Max & 69.07 & 37.83 & 40.94 & 73.67 & 40.08 & 48.05 & 66.73 & 39.13 & 53.16 \\
\quad DeepSeek-V3.2 & 41.24 & 23.33 & 23.61 & 49.65 & 29.11 & 33.53 & 36.05 & 20.38 & 24.14 \\
\quad GPT-5.2 & 50.32 & 30.07 & 30.83 & 46.78 & 27.67 & 32.43 & 62.64 & 39.35 & 43.19 \\
\bottomrule
\end{tabular}
}
\end{table}


To better understand why \oursys successfully translates queries where baselines fail, we further conduct finer-grained analysis in both dialect coverage and LLM backbones. 

\hi{Dialect Coverage.} Executability alone does not guarantee that queries are written idiomatically. \oursys achieves a high DFC across all databases (e.g., 90.07\% on SQLite and 88.42\% on MySQL). This indicates that our system does not merely generate generic SQL to bypass syntax errors, but effectively retrieves and synthesizes native dialect features to systematically overcome complex dialectal conflicts. For instance, \oursys successfully avoids \textit{Unsupported Syntax} by accurately synthesizing Oracle's native \texttt{LISTAGG} instead of hallucinating \texttt{GROUP\_CONCAT}, prevents \textit{Incorrect Usage} by strictly adhering to Oracle's two-argument \texttt{CONCAT} signature, and resolves \textit{Implicit Constraints} by restructuring illegal nested aggregations in MySQL into compliant \texttt{CASE WHEN} constructs.


\noindent \textbf{LLM Backbones.}
We evaluate the performance of \oursys and DIN-SQL across three different LLMs: Qwen-3-Max, DeepSeek-V3.2, and GPT-5.2.
As shown in Table \ref{tab:llm_comparison}, \oursys maintains stable and high performance regardless of the underlying \llm.
For instance, on PostgreSQL, \oursys achieves an execution accuracy between 43.10\% and 53.33\% across the three models. In contrast, DIN-SQL exhibits high variance, dropping from 37.83\% with Qwen-3-Max to 23.33\% with DeepSeek-V3.2. This demonstrates that by explicitly decoupling semantic logic from dialect syntax and relying on an external knowledge base, \oursys significantly reduces the dependency on the \llm's internal (often flawed) parametric knowledge.

\subsection{Ablation Study}
\label{ssec:ablation}

We conduct ablation studies to verify the effectiveness of the three core components of \oursys. The results on PostgreSQL, SQL Server, and Oracle are summarized in Table \ref{tab:ablation_updated}.

Note that AIDE without HINT-KB is undefined because its rule-retrieval step requires $\mathcal{R}_{Rule}$; removing the knowledge base would reduce correction to unconstrained LLM self-repair, which we empirically observe frequently induces severe semantic drift. Therefore, AIDE's execution is inextricably linked to HINT-KB's presence.

\subsubsection{Effectiveness of the Dialect-Aware Logical Query Planning}\label{subsec:effective}
Removing logical planning (Row 3 in Table \ref{tab:ablation_updated}) lowers PostgreSQL Acc from 53.33\% to 38.59\% and DFC from 78.70\% to 53.35\%, confirming the value of separating semantic decomposition from syntax generation. Furthermore, a single-pass NL-LQP reduces DFC from 78.70\%/85.70\%/76.97\% to 68.15\%/69.49\%/66.51\% on PostgreSQL/SQL Server/Oracle, showing that explicit decomposition plus implicit-logic mining captures intermediate operations more reliably.

\subsubsection{Effectiveness of Hierarchical Dialect Knowledge Base}
The HINT-KB component bridges the gap between abstract intents and concrete syntax. When the system operates without the complete hierarchical knowledge base (Row 5), it struggles to identify the correct dialect-specific functions. As a result, the model falls back on its pre-trained biases, causing the Dialect Feature Coverage (DFC) on Oracle to drop from 76.97\% to 48.52\%. This confirms that relying solely on the LLM's internal knowledge or raw documentation is insufficient; the structured, intent-aware mapping provided by HINT-KB is essential for native syntax synthesis.

\subsubsection{Effectiveness of Adaptive \& Iterative Debugging and Evaluation}
We assess the contribution of the iterative correction mechanism by disabling the feedback loop (Row 4). Without execution-driven correction, the execution accuracy on SQL Server decreases from 51.71\% to 44.49\%. Zero-shot generation, even with an accurate knowledge base, cannot anticipate all implicit engine constraints (e.g., transient type-casting rules or specific reserved keyword conflicts). The adaptive debugging mechanism provides critical on-the-fly repairs, ensuring that minor syntactic violations are resolved without causing semantic drift from the original user intent.

\subsection{Case Study}
\label{ssec:study}

We further analyze representative dialect-specific NL2SQL errors in Table~\ref{tab:dialect_errors}. For instance, for the \emph{Unsupported Syntax} cases, baselines often transfer functions from dominant dialects to incompatible engines, such as using MySQL-style \texttt{GROUP\_CONCAT} for Oracle string aggregation (U1), \texttt{date()} instead of Oracle \texttt{TO\_DATE} (U3), or unsupported type casts such as \texttt{AS TEXT} in MySQL (U2). Dialect-SQL also exposes an abstraction leak when ORM bindings fail to map DuckDB native numeric types (U4). In contrast, \oursys separates abstract user intent from concrete dialect realization and retrieves the corresponding native implementation from HINT-KB, such as \texttt{LISTAGG} for Oracle and dialect-valid casting or numeric handling for MySQL and DuckDB.

\section{Related Work}
\label{related_work}

\hi{General NL2SQL.} 
LLM-based NL2SQL has been extensively surveyed~\cite{DBLP:journals/tkde/LiuSLMJZFLTL25, DBLP:journals/pvldb/LiLCLT24, DBLP:journals/pvldb/LuoLFCT25}. 
Representative methods include modular prompting~\cite{DIN-SQL,DAIL-SQL,Pourreza2024@chasesql}, supervised fine-tuning~\cite{dts-sql,codes}, search and feedback~\cite{DBLP:conf/icml/Li0FXC0L25,DBLP:journals/corr/abs-2510-17586}, reward-guided reasoning~\cite{zhang2025rewardsqlboostingtexttosqlstepwise} and complexity-aware routing~\cite{zhu2025elliesql}. 
\emph{However, these methods mainly target a single SQL dialect and do not explicitly decouple semantic planning from dialect-specific realization.}

\hi{Dialect-Specific NL2SQL.} 
Dialect-SQL~\cite{dialect-sql} bridges dialect gaps through an ORM-based intermediate layer, but its portability-oriented abstraction may replace native operators with verbose generic constructs. 
SQL-GEN~\cite{sql-gen} and ExeSQL~\cite{exesql} mitigate dialect data scarcity through synthetic tutorials or execution supervision, while still relying on data-driven internalization of dialect knowledge. 
Dial grounds dialect realization in explicit logical plans, structured dialect knowledge, and execution-verified correction.

\hi{SQL Dialect Translation.} 
Rule-based tools, including SQLGlot~\cite{sqlglot}, jOOQ~\cite{jooq}, and SQLines~\cite{sqlines}, depend on manually maintained translation rules and remain limited on complex queries. 
Recent SQL-to-SQL systems~\cite{zhou2025cracking, CrackSQLDemo, zhou2026dbcooker, zhoumemorydata2026, DBAIOps} further combine LLMs with functionality-based query processing for cross-dialect translation. 
\RThreeMargin{R3.W5}{\RthreeColor{\llabel{R3.W5.cracksql}Unlike NL2SQL, SQL-to-SQL translation assumes complete source-query semantics and focuses on preserving functionality while rewriting dialect-specific fragments.}}

\section{Conclusion}

In this paper, we proposed a dialect-specific NL2SQL framework. 
We introduced {Dialect-Aware Logical Query Planning}, which constructs a {Natural Language Logical Query Plan} (NL-LQP) to decouple semantic intent from dialect-specific syntax. 
We built {HINT-KB}, a hierarchical intent-aware knowledge base that organizes vendor documentation into declarative function mappings and procedural constraint rules to guide generation. 
We further designed an {Adaptive \& Iterative Debugging and Evaluation} mechanism that leverages execution feedback for syntactic recovery while verifying consistency with the logical plan. \RTwoMargin{R2.D4}{}\llabel{R2.D4.general}\RtwoColor{Beyond dialect resiliency, the explicit intent representation of NL-LQP and the objective semantic drift analysis provide a generalized paradigm for intent preservation in broader NL2SQL tasks.}
Experiments on the {DS-NL2SQL} benchmark show that \oursys significantly improves execution accuracy and dialect feature coverage over state-of-the-art baselines.

There are several potential directions. 
First, we will incorporate lightweight dialect parsers to reduce reliance on live database feedback. Second, we will improve knowledge acquisition to support niche dialects or legacy databases lacking  documentation.

\section*{Acknowledgment} Xuanhe Zhou, Wei Zhou, and Fan Wu are the corresponding authors. 
This paper was supported in part by Fundamental and Interdisciplinary Disciplines Breakthrough Plan of the Ministry of Education of China (No. JYB2025XDXM103), in part by National Key R\&D Program of China (2023YFB4503600), NSF of China (62432007, 62441236, 62372296, 62525202, 62232009, 62502304, U25A20437), in part by Huawei, Tencent, ByteDance, Shanghai Artificial Intelligence Laboratory.

\clearpage

\bibliographystyle{ACM-Reference-Format}
\bibliography{refs}

\end{document}